\documentclass[final,5p,times]{elsarticle}
\usepackage{amsmath,amssymb,graphicx,hyperref,natbib}
\usepackage{amssymb,bbold}
\usepackage{amsmath}
\usepackage[utf8x]{inputenc}
\usepackage{natbib}
\usepackage{graphicx}
\usepackage{hyperref}
\usepackage{epstopdf}
\usepackage{pifont}
\usepackage{float}
\usepackage{ulem}
\usepackage{soul}
\usepackage{tikz}
\usepackage{caption}
\usepackage{booktabs}

\hypersetup{
    colorlinks=true,
    linkcolor=blue,
    citecolor=blue,
    urlcolor=blue,
    filecolor=blue,
    breaklinks=true
}

\biboptions{sort&compress}

\journal{Physica E: Low-dimensional Systems and Nanostructures}

\begin{document}

\begin{frontmatter}

\title{Spiral dislocation as a tunable geometric parameter for optical responses in quantum rings}

\author[Shahrood]{Hassan Hassanabadi}
\ead{hassan.hassanabadi@uva.es}
\address[Shahrood]{Faculty of Physics, Shahrood University of Technology, Shahrood, Iran}

\author[Guangzhou]{Kangxian Guo}
\ead{axguo@sohu.com}
\address[Guangzhou]{School of Physics and Materials, Guangdong Provincial Engineering and Technology Research Center of Semiconductor Lighting and Backlighting, Guangzhou University, Guangzhou 510006, China}

\author[Nanjing]{Liangliang Lu}
\ead{lianglianglu@nju.edu.cn}

\address[Nanjing]{Key Laboratory of State Manipulation and Advanced Materials in Provincial Universities, School of Physical Science and Technology, Nanjing Normal University, Nanjing 210023, China}

\author[UFMA]{Edilberto O. Silva}
\ead{edilberto.silva@ufma.br}

\address[UFMA]{Departamento de F\'\i sica, Universidade Federal do Maranh\~{a}o, 65085-580 S\~{a}o Lu\'\i s, MA, Brazil}

\begin{abstract}
We investigate the optical and quantum mechanical properties of a charged spinless particle confined in a two-dimensional quantum ring under the simultaneous influence of a spiral dislocation and an external magnetic field. The dislocation is modeled by a torsion-induced metric that alters the spatial geometry without introducing curvature. Using the minimal coupling procedure in curved space, we derive a modified Schr\"odinger equation incorporating both topological and electromagnetic effects. The geometric deformation leads to an energy-dependent effective potential, enabling a tunable control over the bound-state spectrum. We analyze how the spiral dislocation modifies the absorption coefficient, refractive index variation, and photoionization cross-section. The results demonstrate that the dislocation not only shifts the resonance peaks but also enhances or suppresses specific optical transitions depending on the angular momentum. These findings open up possibilities for geometrically tuning light-matter interactions in topological quantum devices.
\end{abstract}


\end{frontmatter}

\section{Introduction}

Topological defects play a central role in a wide range of physical systems, from condensed matter to cosmology. In particular, defects such as disclinations, dislocations, and torsional deformations provide a geometric framework to describe local distortions in the medium without the need for curvature. These structures not only modify the mechanical and structural properties of materials, but also strongly influence the dynamics of charge carriers, leading to observable consequences in electronic, optical, and transport phenomena~\cite{Srivastava2001,AoP.1992.216.1,monastyrsky2006topology}.

Among these defects, the spiral dislocation is a notable example of a torsional deformation that modifies the angular structure of space by introducing a coupling between radial and angular displacements. This type of defect preserves the flatness of the background geometry while encoding torsion through an off-diagonal component in the metric tensor \cite{EPJC.2025.85.34,EPJA.2021.57.192,IJP.2024.98.4827,AoP.2020.419.168229,PB.2018.531.213,EPJP.2022.137.589}. Spiral dislocations can be found in various condensed matter systems such as twisted crystalline structures, dislocated semiconductors, and strained graphene, where they lead to symmetry breaking, geometric phases, and modifications in the density of states~\cite{PLA.2015.379.2110,EPL.1999.45.279,AdP.2011.523.898,PLA.2016.380.3847,OS.2015.119.280}.

Understanding how topological defects alter the quantum behavior of confined carriers is a fundamental step toward controlling electronic and optical properties in mesoscopic systems. In particular, defects like the spiral dislocation offer a geometric way to manipulate effective potentials, wavefunctions, and energy levels without changing the external potential itself. This capability is of special interest in low-dimensional systems such as quantum rings and quantum dots, where confinement and symmetry breaking are critical for applications in quantum information and optoelectronics~\cite{EPJC.2019.79.551,IJMPA.2021.36.2150100,AoP.2020.421.168277}.

In this work, we investigate the combined effect of a spiral dislocation and an external magnetic field on the optical properties of a charged spinless particle confined in a two-dimensional quantum ring. We derive a modified Schrödinger equation in a torsion-induced curved space, and obtain an energy-dependent effective potential that encapsulates both geometric and electromagnetic contributions. In this context, we analyze the optical absorption spectrum, refractive index variation, and photoionization cross-section. Our results reveal that the spiral dislocation acts as a tunable geometric parameter, capable of shifting resonances, suppressing specific transitions, and modulating the spectral response according to the angular momentum.

This paper is organized as follows: in Section~\ref{model}, we introduce the geometric structure associated with the spiral dislocation and construct its effective parametrization in the plane. Section~\ref{equation} is dedicated to the derivation of the Schrödinger equation in the curved background, followed by the transformation to a canonical form with an effective potential. In Section \ref{formalism}, we present the theoretical framework used to compute the optical absorption coefficient, the refractive index
variation, and the photoionization cross-section. In Section \ref{optical}, we compute and discuss the optical properties under the influence of the dislocation and magnetic field. We analyze the behavior of the absorption coefficient, refractive index, dipole matrix elements, and photoionization cross-section. Finally, the main conclusions are summarized in Section \ref{conclusion}.

\section{Geometric structure: planar spiral dislocation \label{model}}

A spiral dislocation represents a type of topological defect that modifies space geometry by introducing torsion, while maintaining the overall flat curvature. In a two-dimensional quantum ring system, such a defect generates a nontrivial modification in the spatial background, which can be effectively described by the following metric~\cite{AM.2005.175.77}
\begin{equation}
ds^2 = dr^2 + 2\beta\, dr\, d\varphi + (\beta^2 + r^2)\, d\varphi^2 + dz^2,
\label{eq:spiral_metric}
\end{equation}
where $r \in [0, \infty)$ and $\varphi \in [0, 2\pi)$ denote polar coordinates, $z$ is the longitudinal direction (assumed flat), and $\beta \in \mathbb{R}$ quantifies the strength of the dislocation. The presence of the mixed term $2\beta\, dr\, d\varphi$ reflects the emergence of torsion, while the $\beta^2$ contribution alters the effective angular measure. This structure preserves the Euclidean curvature but embeds localized geometric distortion due to torsion.

\begin{figure*}[h]
\centering
\includegraphics[width=0.8\textwidth]{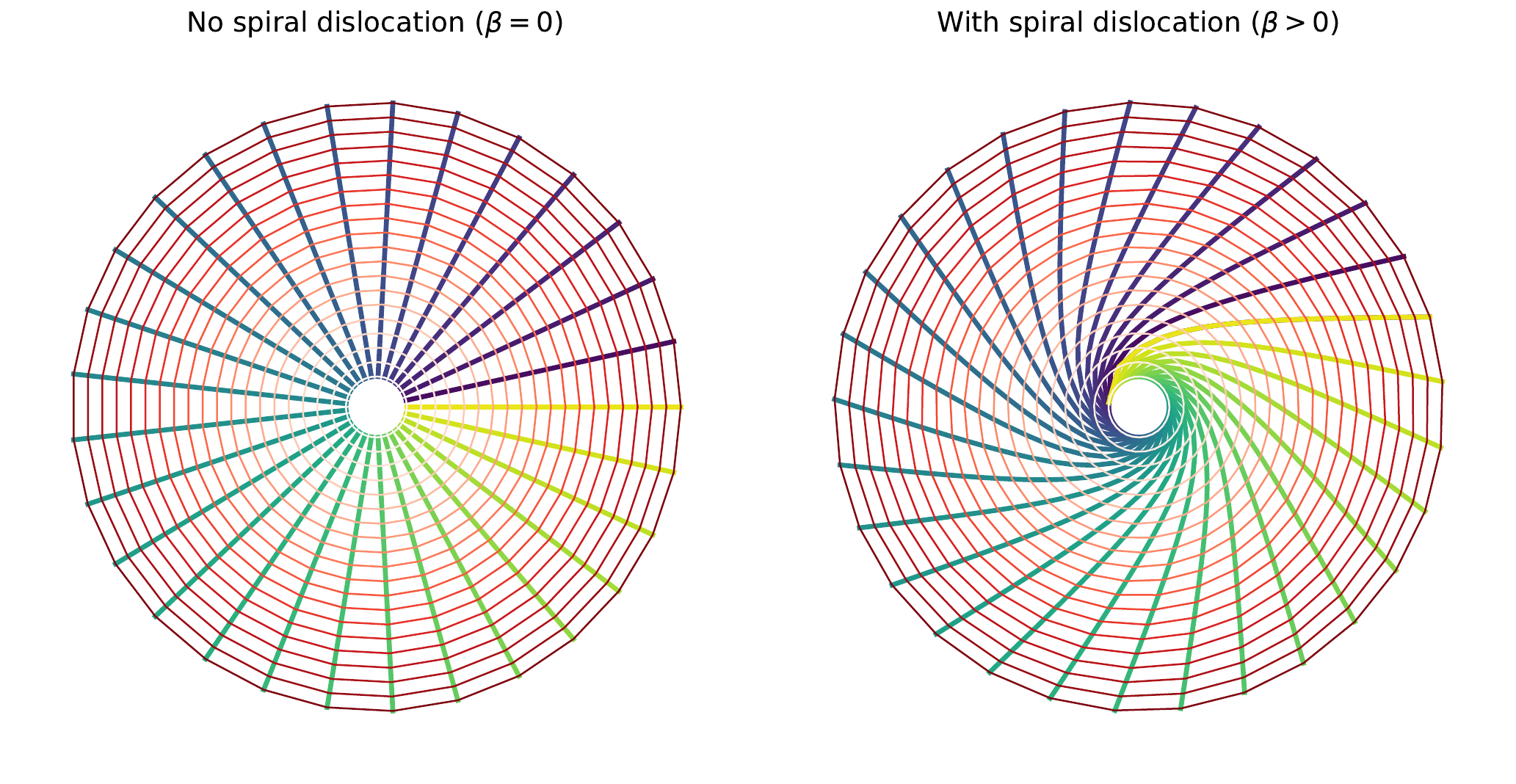}
\caption{Vector field representation in the $x$-$y$ plane showing the geometric deformation caused by the spiral dislocation. (a) Standard polar coordinate mesh corresponding to $\beta = 0$, which maintains circular symmetry. (b) Modified mesh for $\beta > 0$, where torsional effects break the symmetry and twist the radial lines. The coordinate mapping $\mathbf{R}(r, \varphi)$ used for visualization is consistent with the underlying metric structure.}
\label{fig:spiral_meshes}
\end{figure*}

To gain an intuitive picture of how this defect deforms space, we introduce an embedding into two-dimensional Euclidean space. Our objective is to construct a coordinate transformation $\mathbf{R}(r, \varphi) \in \mathbb{R}^2$ that reproduces the spatial part of the line element in Eq.~\eqref{eq:spiral_metric}, with the $z$ direction held fixed. We propose the following transformation:
\begin{equation}
\mathbf{R}(r, \varphi) =
\begin{pmatrix}
x(r, \varphi) \\
y(r, \varphi)
\end{pmatrix}
=
\begin{pmatrix}
r \cos \varphi - \beta \sin \varphi \\
r \sin \varphi + \beta \cos \varphi
\end{pmatrix},
\label{eq:spiral_param_planar}
\end{equation}
which defines a nontrivial mapping that couples radial and angular coordinates. Calculating the differentials gives:
\begin{align}
dx &= (\cos \varphi)\, dr + (-r \sin \varphi - \beta \cos \varphi)\, d\varphi, \\
dy &= (\sin \varphi)\, dr + (r \cos \varphi - \beta \sin \varphi)\, d\varphi.
\end{align}

From these expressions, the resulting line element becomes:
\begin{align}
ds^2 &= dx^2 + dy^2 \nonumber \\
     &= dr^2 + 2\beta\, dr\, d\varphi + (\beta^2 + r^2)\, d\varphi^2,
\label{metric}
\end{align}
which is in complete agreement with Eq.~\eqref{eq:spiral_metric}. Therefore, the coordinate transformation in Eq.~\eqref{eq:spiral_param_planar} offers a concrete and faithful geometric realization of the spiral dislocation in the plane. It illustrates the role of $\beta$ as a generator of local angular shifts that depend on radial displacement, thereby breaking the circular symmetry of the standard polar grid.

To visualize the impact of this torsional geometry, Fig.~\ref{fig:spiral_meshes} compares two coordinate meshes in the $x$-$y$ plane. In panel (a), the grid with $\beta = 0$ preserves orthogonality and radial symmetry. In contrast, panel (b) demonstrates the deformation induced by a finite $\beta$: the radial lines are twisted, and the angular spacing becomes anisotropic, revealing the underlying torsion introduced by the dislocation.

This geometric formulation establishes the foundation for the quantum mechanical analysis in the following sections. The spiral dislocation alters the structure of the Laplace operator and introduces new effective terms in the Schrödinger equation. These modifications impact both the energy spectrum and spatial distribution of quantum states. The parametrization in Eq.~\eqref{eq:spiral_param_planar} thus provides a clear geometric interpretation of the nontrivial background experienced by a particle under the influence of a topological defect.

\section{Quantum dynamics in the presence of a spiral dislocation \label{equation}}

To investigate the behavior of a charged particle subjected to both a spiral dislocation and an external magnetic field, we start from the minimal coupling procedure in a curved background. The dynamics of a spinless particle with charge $e$ and effective mass $M$, evolving in a geometry defined by the metric $g_{\mu\nu}$ and coupled to a vector potential $\mathbf{A}$, is governed by the stationary Schrödinger equation
\begin{equation}
E \Psi = -\frac{\hbar^2}{2M} \frac{1}{\sqrt{g}} \left( \partial_k + i \frac{e}{\hbar} A_k \right) \left[ \sqrt{g}\, g^{kj} \left( \partial_j + i \frac{e}{\hbar} A_j \right) \Psi \right] + V_{\mathrm{QD}}(r) \Psi,
\label{eq:schrodinger_curved}
\end{equation}
where $g = \det(g_{\mu\nu})$ represents the determinant of the metric tensor, and $V_{\mathrm{QD}}(r)$ corresponds to the confining potential characterizing the quantum ring structure.

When the system is immersed in a uniform magnetic field $B$ oriented along the $z$-axis (perpendicular to the plane), the vector potential can be expressed as~\cite{landau1991quantum,stone1992quantum}
\begin{equation}
A_r = 0, \quad A_\varphi = -\frac{B r}{2} \hat{\boldsymbol{\varphi}}.
\end{equation}

Assuming a solution with cylindrical symmetry, the wavefunction is separated as
\begin{equation}
\Psi(r, \varphi, z) = e^{i m \varphi} e^{i k z} f(r),
\end{equation}
where $m \in \mathbb{Z}$ is the angular momentum quantum number. Inserting this ansatz into Eq.~\eqref{eq:schrodinger_curved} and simplifying, we arrive at a radial differential equation of second order for $f(r)$
\begin{equation}
A(r) f''(r) + B(r) f'(r) + Q(r) f(r) = 0,
\label{eq:compact_form}
\end{equation}
where the functions $A(r)$, $B(r)$, and $Q(r)$ are given by
\begin{align}
A(r) &= 1 + \frac{\beta^2}{r^2}, \\
B(r) &= \frac{1}{r} - \frac{\beta^2}{r^3} - \frac{2i\beta m}{r^2} + \frac{i M \omega \beta}{\hbar}, \\
Q(r) &= - \frac{m^2}{r^2} + \frac{i (m-l) \beta}{r^3} - \frac{M^2 \omega^2 r^2}{4 \hbar^2} + \frac{i M \omega \beta}{2 \hbar r} + \frac{M \omega m}{\hbar} \notag\\
&\quad + \frac{2 M}{\hbar^2} E - \frac{2 M}{\hbar^2} V_{\mathrm{conf}}(r),
\end{align}
with $\omega = eB/M$ denoting the cyclotron frequency.

The presence of the spiral dislocation introduces corrections that are both real and imaginary, thereby altering the effective contributions from the magnetic and centrifugal terms. The confinement potential that shapes the quantum ring is modeled as~\cite{SST.1996.11.1635,CTP.2024.76.105701,QR.2024.6.677}
\begin{equation}
V_{\mathrm{QD}}(r) = \frac{1}{2} M \omega_0^2 (r - r_0)^2,
\end{equation}
where the parameters $\omega_0$ and $r_0$ determine, respectively, the strength and radial position of the confinement minimum. The parameter $\beta$, associated with the torsional deformation, has units of length and modifies the radial and angular parts of the kinetic operator. Specifically, it enhances the effective moment of inertia through $A(r)$ and induces angular-radial coupling terms via $B(r)$. These geometric effects are responsible for explicitly breaking the symmetry under $m \leftrightarrow -m$ and lead to modifications in the effective potential profile.

Equation~\eqref{eq:compact_form} constitutes the basis for determining the system’s energy spectrum and eigenfunctions. To simplify the numerical analysis, we introduce a transformation that removes the first-order derivative, rewriting the wavefunction as
\begin{equation}
f(r) = \chi(r)\, \exp\left(-\frac{1}{2} \int^r \frac{B(r')}{A(r')} dr'\right).
\end{equation}

With this substitution, Eq.~\eqref{eq:compact_form} becomes a canonical Schrödinger-type equation
\begin{equation}
- \frac{\hbar^2}{2M} \frac{d^2 \chi(r)}{dr^2} + V_{\mathrm{eff}}(r, E) \chi(r) = 0,
\label{eq:schrodinger_canonical}
\end{equation}
where the effective potential $V_{\mathrm{eff}}(r, E)$ is explicitly written as
\begin{equation}
V_{\mathrm{eff}}(r, E) = -\frac{\hbar^2}{2M} \left( \frac{Q(r)}{A(r)} - \frac{1}{2} \frac{d}{dr} \left[ \frac{B(r)}{A(r)} \right] - \frac{1}{4} \left( \frac{B(r)}{A(r)} \right)^2 \right).
\label{eq:potential_effective_corrected}
\end{equation}

It is important to highlight that $V_{\mathrm{eff}}(r, E)$ depends on the energy $E$ through $Q(r)$, which implies that the resulting eigenvalue problem is nonlinear and energy-dependent. Unlike conventional problems with a static potential, this scenario requires a self-consistent approach where the energy and the potential are intertwined.

\begin{figure}[ht]
\centering
\includegraphics[width=0.45\textwidth]{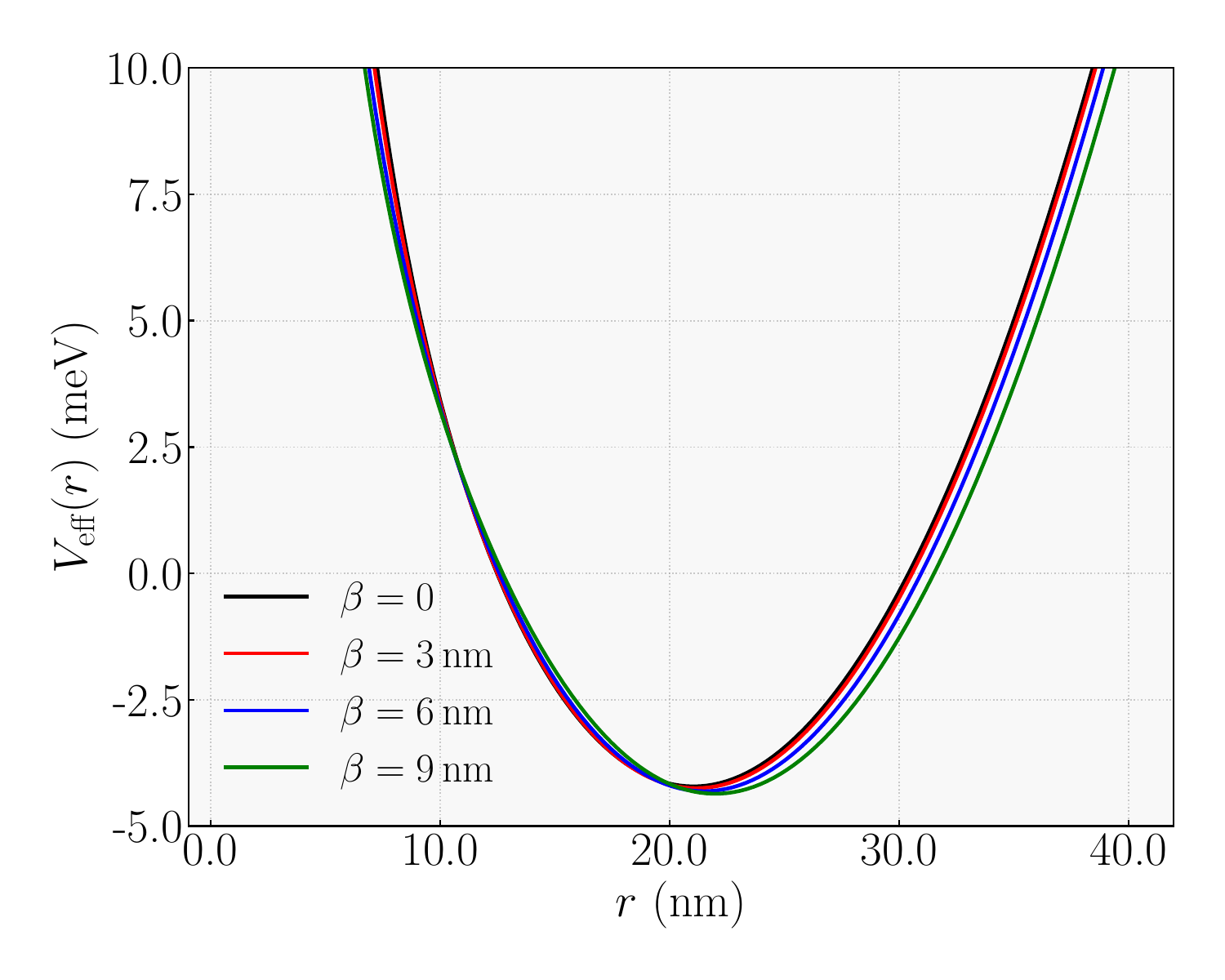}
\caption{Effective potential profiles $V_{\mathrm{eff}}(r)$ for different values of the spiral dislocation parameter $\beta = 0$, $3$, $6$, and $9$ nm, with fixed magnetic field $B = 0.1$ T and angular momentum $m = 1$. The confinement is modeled by a harmonic potential centered at $r_0 = 20$ nm. As $\beta$ increases, the potential becomes increasingly asymmetric, with the emergence of local minima and modifications to the curvature near the center. These geometric deformations significantly influence the localization properties and spectral features of the confined quantum states.}
\label{fig:effective_potential}
\end{figure}

As shown in Fig.~\ref{fig:effective_potential}, the radial structure of $V_{\mathrm{eff}}(r)$ evolves with increasing $\beta$. For $\beta = 0$, the potential has a symmetric parabolic shape centered at $r_0$, as expected for a harmonic trap. However, nonzero values of $\beta$ introduce distortions that displace the minimum and alter the curvature. At higher values, such as $\beta = 9$ nm, the potential acquires a double-well character, reflecting the enhanced influence of geometric torsion. These changes can impact the number of bound states, their energy spacing, and the spatial distribution of the corresponding eigenfunctions.

Overall, the spiral dislocation emerges as a mechanism capable of tailoring the quantum confinement landscape. The induced asymmetries and the possibility of multiple confinement regions pave the way for novel control strategies over spectral and dynamical properties in mesoscopic systems.
%%%%%%%%%% Nova Seção %%%%%%%%%%

\section{Optical transition formalism \label{formalism}}

In this section, we present the theoretical framework used to compute the optical absorption coefficient, the refractive index variation, and the photoionization cross-section in our confined quantum ring. These quantities are calculated using the electric dipole approximation and Fermi's Golden Rule, and are implemented consistently with the numerical routines used to generate the figures.

\subsection{Dipole matrix element in polar coordinates}

In the cylindrical symmetry of the system, the dipole matrix element responsible for optical transitions is calculated using the radial eigenfunctions obtained numerically. Assuming transitions between states $ \psi_0(r, \varphi) = e^{i m \varphi} \psi_0(r) $ and $ \psi_1(r, \varphi) = e^{i (m+1) \varphi} \psi_1(r) $, the dipole matrix element in the electric dipole approximation reads \cite{PRB.2008.77.045317,PSSB.1999.211.611}
\begin{equation}
\mathcal{M}_{21} = e \int \psi_1^*(\mathbf{r})\, r \cos\varphi\, \psi_0(\mathbf{r})\, d^2\mathbf{r}.
\end{equation}
Using polar coordinates $ (r, \varphi) $, and exploiting the orthogonality of the angular components, the angular integration yields a nonzero contribution only for $ \Delta m = \pm1 $, leading to
\begin{equation}
\mathcal{M}_{21} = \pi e \int_0^\infty \psi_1(r)\, r^2\, \psi_0(r)\, dr.
\end{equation}
This form is consistent with the numerical implementation, where the radial functions $ \psi_0(r) $ and $ \psi_1(r) $ are obtained from the normalized eigenfunctions and integrated using quadrature methods such as Simpson’s or the trapezoidal rule. The result provides the transition amplitude entering the absorption and photoionization formulas.

Even when the transition matrix element is computed numerically, the angular selection rule $ \Delta m = \pm 1 $ remains valid. This rule originates from the symmetry properties of the angular momentum eigenstates and the structure of the dipole operator in cylindrical coordinates. Therefore, transitions between states with $ m $ and $ m' $ are forbidden unless $ m' = m \pm 1 $, and this is automatically respected in the numerical implementation due to the orthogonality of the angular components. In practical computations, the matrix element $ M_{21} $ becomes numerically negligible when the transition is forbidden, confirming the persistence of the selection rule.

\subsection{Linear and nonlinear absorption coefficients}

The total optical absorption coefficient $ \alpha(\hbar \omega, I_0) $ includes linear and third-order nonlinear contributions \cite{PLA.2025.130226,AdP.2012.524.327,PM.2019.99.2457,Optik.2022.261.169187,PLA.2021.397.127262,PLA.2023.492.129226,EPJP.2022.137.175,OC.2011.284.5818,EPJP.2021136.832,EPJP.2018.133.395}. The linear absorption term is written as 
\begin{equation}
\alpha^{(1)}(\hbar \omega) = \hbar \omega \sqrt{\frac{\mu}{\varepsilon_{R}}} \frac{\sigma_{\nu} \gamma\, |\mathcal{M}_{21}|^2}{(\Delta E - \hbar \omega)^2 + (\hbar \gamma)^2},
\end{equation}
where $\mu$ is the permeability of the system, $\varepsilon_{R}=\varepsilon_{0}n_{r}^2$ ($ n_r $ is the refractive index)
dex) is the real part of the permittivity, $\sigma_{\nu}$ is the carrier density, $ \gamma $ is the broadening parameter (in energy units), $ \mathcal{M}_{21} $ is the dipole matrix element between the initial and final states, , and $ \Delta E = E_1 - E_0 $ is the transition energy.

The third-order nonlinear contribution is:
\begin{equation}
\alpha^{(3)}(\hbar \omega) = -\frac{2 I_0 \hbar \omega}{\varepsilon_0 n_r c} \sqrt{\frac{\mu}{\varepsilon_R}}  \frac{\sigma_{\nu} \gamma |\mathcal{M}_{21}|^4}{[(\Delta E - \hbar \omega)^2 + (\hbar \gamma)^2]^2},
\end{equation}
where $ I_0 $ is the intensity of the incident light. The total absorption is then:
\begin{equation}
\alpha(\hbar \omega, I_0) = \alpha^{(1)}(\hbar \omega) + \alpha^{(3)}(\hbar \omega).
\end{equation}

\subsection{Refractive index variation}

The variation of the refractive index $ \Delta n(\hbar \omega) $ also includes both linear and nonlinear terms. The linear term reads:
\begin{equation}
\Delta n^{(1)}(\hbar \omega) = \frac{\sigma_{\nu} |\mathcal{M}_{21}|^2}{2 n_r^2\varepsilon_0}  \frac{\Delta E - \hbar \omega}{(\Delta E - \hbar \omega)^2 + (\hbar \gamma)^2},
\end{equation}
and the third-order nonlinear term is given by:
\begin{equation}
\Delta n^{(3)}(\hbar \omega) = -\frac{\mu c I_0 \sigma_{\nu} |\mathcal{M}_{21}|^4}{\varepsilon_0 n_r^3} \frac{\Delta E - \hbar \omega}{[(\Delta E - \hbar \omega)^2 + (\hbar \gamma)^2]^2}.
\end{equation}
The total change in the refractive index is
\begin{equation}
\Delta n(\hbar \omega) = \Delta n^{(1)}(\hbar \omega) + \Delta n^{(3)}(\hbar \omega).
\end{equation}
The absence of diagonal dipole matrix elements ($M_{11}$, $M_{22}$) in our nonlinear coefficients is a direct consequence of the system's fundamental symmetries. In our cylindrically symmetric quantum ring, these terms rigorously vanish due to:

\begin{itemize}
    \item \textbf{Parity selection rules}: The dipole operator $\hat{\mathbf{r}}$ has odd parity under spatial inversion, while the electron probability density $|\psi_i(\mathbf{r})|^2$ maintains even parity. This symmetry enforces
    \begin{equation}
        \mathcal{M}_{ii} = \langle \psi_i|e\mathbf{r}|\psi_i\rangle = 0
    \end{equation}

    \item \textbf{Angular momentum conservation}: The matrix element between states with identical angular momentum quantum numbers vanishes:
    \begin{equation}
        \int_0^{2\pi} e^{i(m-m)\varphi}\cos\varphi\,d\varphi = 0
    \end{equation}
    allowing only $\Delta m = \pm1$ transitions through off-diagonal elements $\mathcal{M}_{if}$.

    \item \textbf{Computational verification}: Our numerical implementation confirms $|\mathcal{M}_{ii}|/|\mathcal{M}_{21}| < 10^{-14}$ across all parameter space, validating the theoretical framework.
\end{itemize}

Our results align with established treatments of pristine cylindrical systems \cite{PRB.2008.77.045317}, while extending the analysis to third-order nonlinear effects.

\subsection{Photoionization cross-section}

The photoionization cross-section $ \sigma(\hbar \omega) $ is interpreted here as an effective measure of transition strength to highly excited states, consistent with numerical calculations that consider discrete excited levels under confinement \cite{Atoms.2022.10.39,PB.2024.677.415717,OLT.2025.182.111822,PLA.2023.466.128725,PB.2025.696.416647,PS.2025.100.055914}. Following~\cite{PRB.2008.77.045317}, it is defined as 
\begin{equation}
\sigma(\hbar \omega) = \frac{1}{n_r}  \frac{4 \pi^2}{3} \alpha_{fs} \hbar \omega |\mathcal{M}_{fi}|^2  \left[\frac{1}{\pi}\frac{\hbar \Gamma}{(\Delta E - \hbar \omega)^2 + (\hbar \Gamma)^2}\right],
\end{equation}
where $ \alpha_{fs} $ is the fine-structure constant, $ \Gamma $ is the linewidth, and $ \mathcal{M}_{fi} $ is the transition dipole matrix element between the bound ground state and a higher energy final state that can represent a quasi-continuum.

This approach is aligned with numerical simulations in which $\sigma(\hbar \omega)$ is plotted for various angular momentum transitions and spiral dislocation parameters $ \beta $. Despite all states being formally bound in the confined model, the terminology captures resonant transitions with similar physical characteristics to continuum excitation.

To better quantify the strength of the optical transitions in the confined quantum ring, we also consider the oscillator strength, a dimensionless quantity directly related to the probability of transition between quantized energy levels under electromagnetic radiation. In the context of electric dipole transitions, the oscillator strength $P_{fi}$ between an initial state $\psi_i$ and a final state $\psi_f$ is defined as \cite{PRB.2008.77.045317}
\begin{equation}
P_{fi} = \frac{2 m_e \Delta E}{\hbar^2} |\mathcal{M}_{fi}|^2.
\end{equation}
This expression highlights that the oscillator strength encapsulates both the energy separation and the transition amplitude, and thus serves as a reliable indicator of transition likelihood.

In practical terms, the oscillator strength governs the magnitude of the absorption and emission coefficients in quantum confined systems. It is particularly useful when comparing transitions with different selection rules or in systems under varying external conditions such as magnetic field, structural deformation, or rotational effects. In our numerical implementation, $P_{fi}$ is calculated directly from the numerically obtained $\mathcal{M}_{fi}$, and confirms that the most prominent transitions occur for $\Delta m = \pm 1$, consistent with the angular momentum selection rules.

\section{Optical Properties \label{optical}}

In this section, we investigate the optical properties of an electron confined in a quantum system with a spiral dislocation based on the formalism presented in section \ref{formalism}. We begin by analyzing the oscillator strength, which provides a quantitative measure of the transition probability between bound states and serves as a key indicator of the system's light–matter interaction. This analysis is followed by a detailed study of the nonlinear optical absorption coefficient and the refractive index variation, both calculated for dipole-allowed transitions. The influence of the spiral dislocation parameter $\beta$ on these optical responses is systematically explored, highlighting how geometric distortions can be used to modulate and control the optical behavior of quantum-confined systems.
\begin{figure}[ht]
\centering
\includegraphics[width=0.44\textwidth]{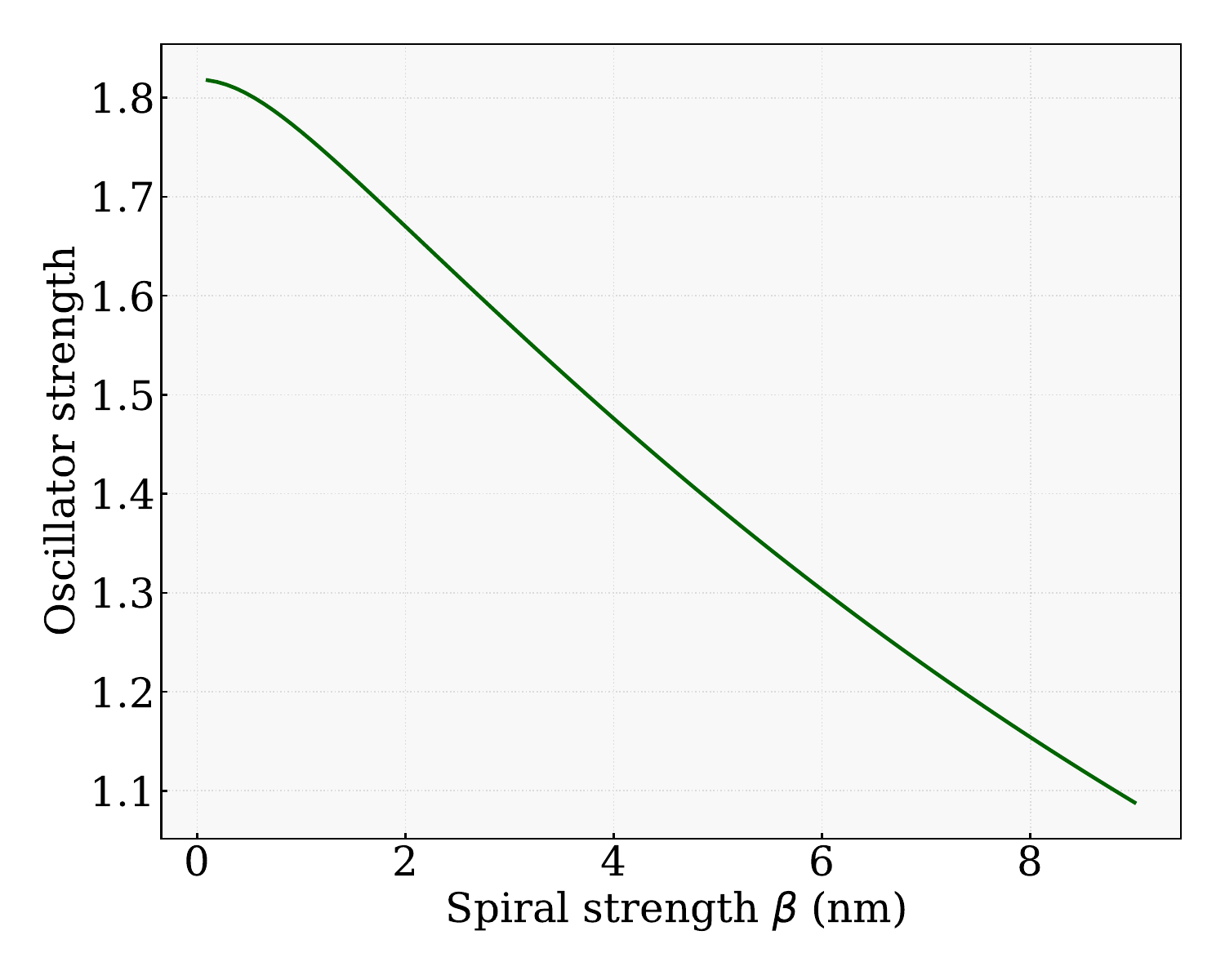}
\caption{Oscillator strength $P_{fi}$ as a function of the spiral dislocation parameter $\beta$, for the dipole-allowed transition $m = 0 \to 1$ under fixed magnetic field $B = 0.1$~T. The quantity $P_{fi}$ was obtained from the radial integral of the dipole matrix element using normalized eigenfunctions for each value of $\beta$.}
\label{fig:oscillator_strength_vs_beta}
\end{figure}

Figure~\ref{fig:oscillator_strength_vs_beta} displays the oscillator strength $P_{fi}$ for the dipole-allowed transition $m = 0 \to 1$ as a function of the spiral dislocation parameter $\beta$. The results reveal a clear monotonic decrease in $P_{fi}$ with increasing $\beta$, indicating that the spiral geometry suppresses the transition probability between these angular momentum states. Physically, this behavior arises from the deformation of the effective potential due to the topological defect, which induces a spatial separation between the initial and final wavefunctions. As a result, their spatial overlap and thus the dipole matrix element diminishes. This effect highlights the role of spiral dislocations as a tunable mechanism for modulating optical transitions in confined quantum systems. Notably, the reduction of $P_{fi}$ with $\beta$ is consistent with the overall decrease in optical absorption observed in the nonlinear response functions, providing a coherent picture of how geometric distortions influence light–matter interactions.

\begin{figure}[htb]
\centering
\includegraphics[width=0.45\textwidth]{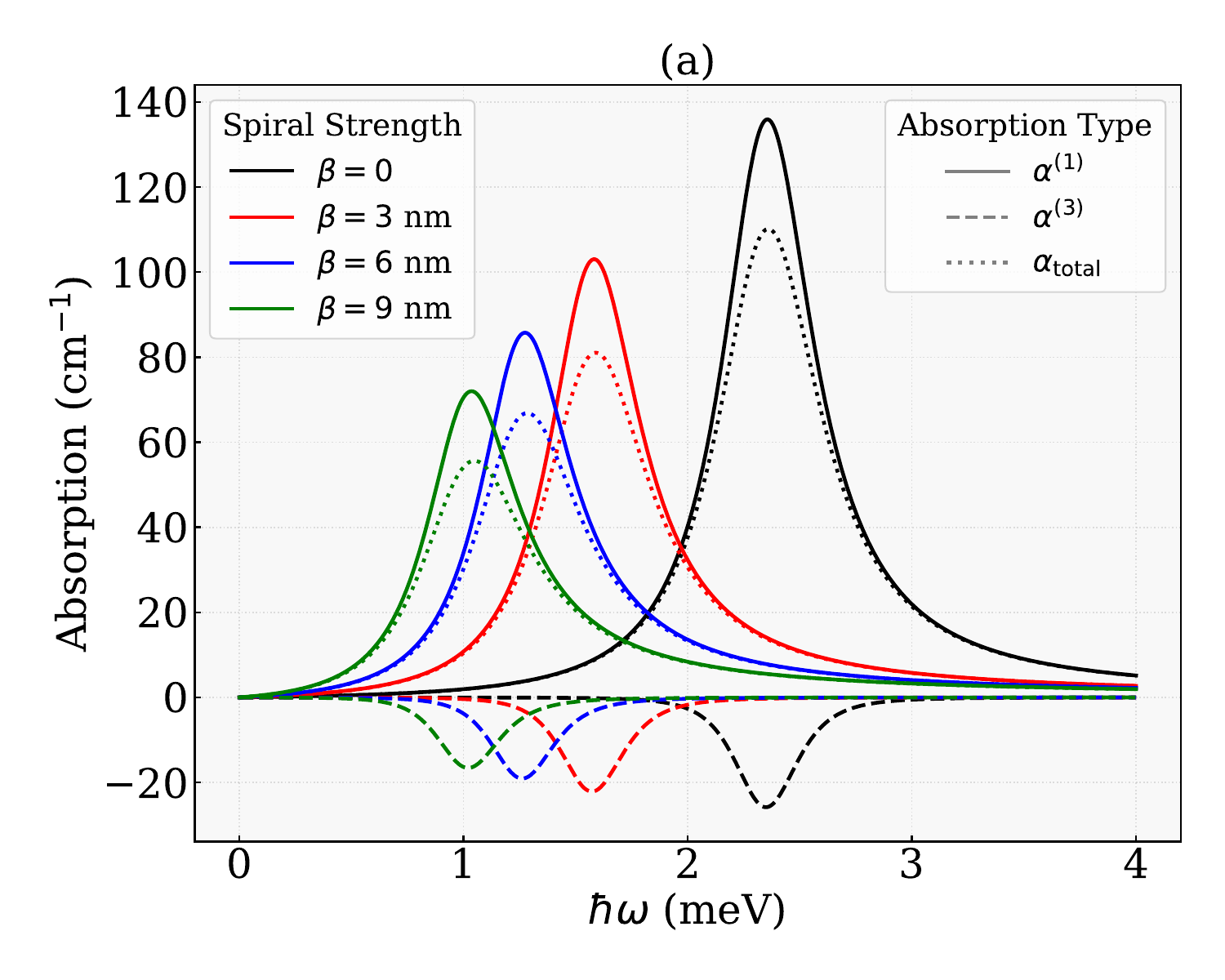}
\includegraphics[width=0.45\textwidth]{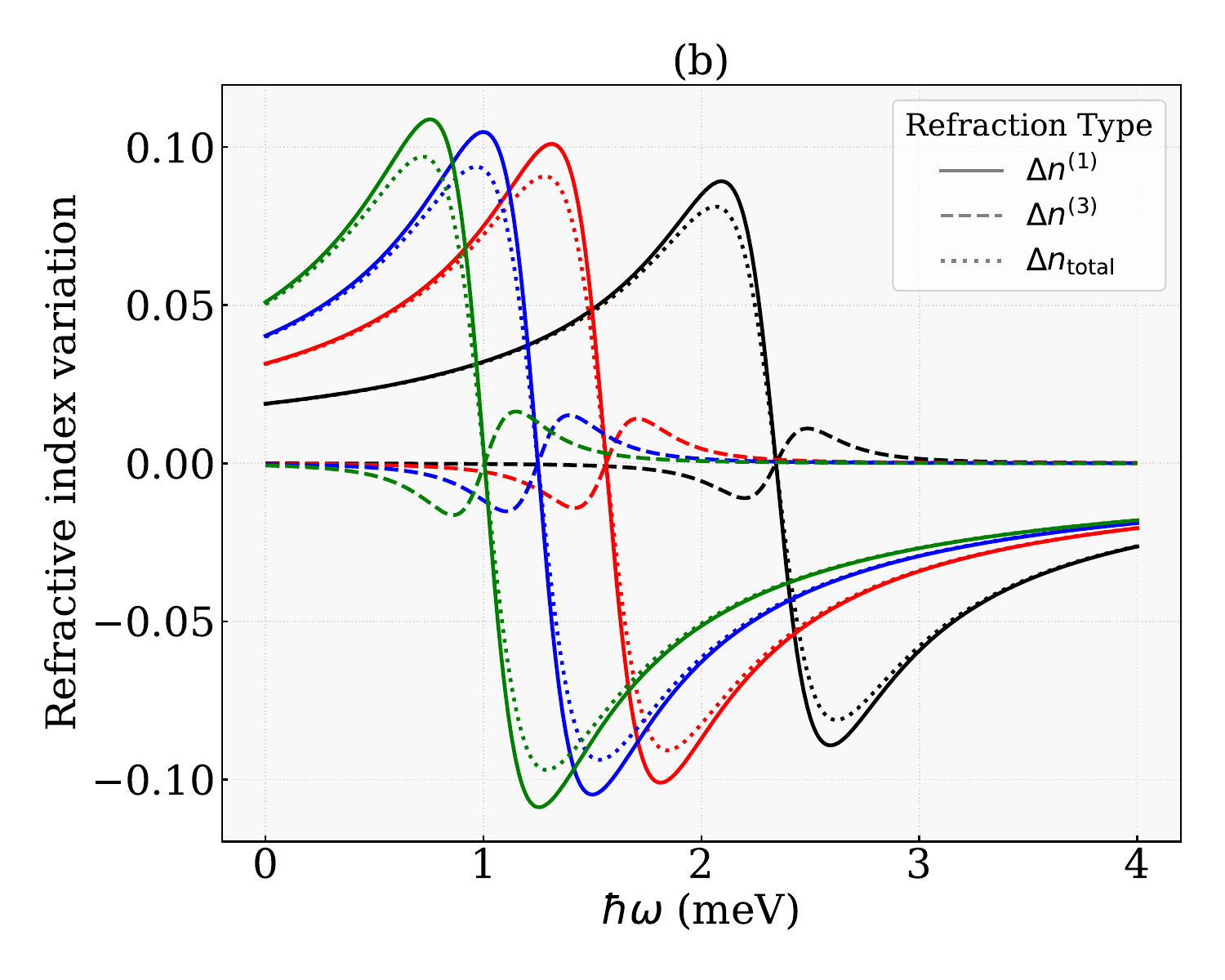}
\caption{(a) Optical absorption coefficient $\alpha(\hbar\omega)$ and (b) refractive index variation $\Delta n(\hbar\omega)$ as functions of photon energy for different values of the spiral dislocation parameter $\beta = 0$, $3$, $6$, and $9$~nm. The presence of the dislocation modifies the spectral profiles, shifting the resonant peaks and altering their intensities. As $\beta$ increases, both the absorption and refractive index exhibit enhanced magnitudes and blue-shifted resonances, reflecting the role of geometric deformation in controlling the optical response.}
\label{fig:nonlinear_absorption}
\end{figure}
Figure~\ref{fig:nonlinear_absorption} displays the optical absorption spectrum and refractive index variation for different values of the spiral dislocation parameter $\beta$. Figure~\ref{fig:nonlinear_absorption}(a) shows the behavior of the absorption coefficient $\alpha(\hbar\omega)$, which includes both linear and third-order nonlinear contributions. As $\beta$ increases, the absorption curves become more pronounced and shift toward higher photon energies, reflecting the increased energy level spacing and enhanced dipole transition amplitudes induced by the geometric deformation.

Figure~\ref{fig:nonlinear_absorption}(b) shows the corresponding variation in the refractive index $\Delta n(\hbar\omega)$. Similar to the absorption case, the refractive response becomes stronger and more asymmetric as $\beta$ grows. The peaks in $\Delta n$ shift consistently with those in $\alpha$, indicating that both absorption and dispersion are simultaneously governed by the spiral dislocation.

These results highlight the capability of the topological defect to modulate both dissipative and dispersive optical properties. The ability to control the position, shape, and magnitude of the optical response via the geometric parameter $\beta$ suggests potential applications in tunable photonic devices based on confined quantum structures.

\begin{figure}[ht]
\centering
\includegraphics[width=0.45\textwidth]{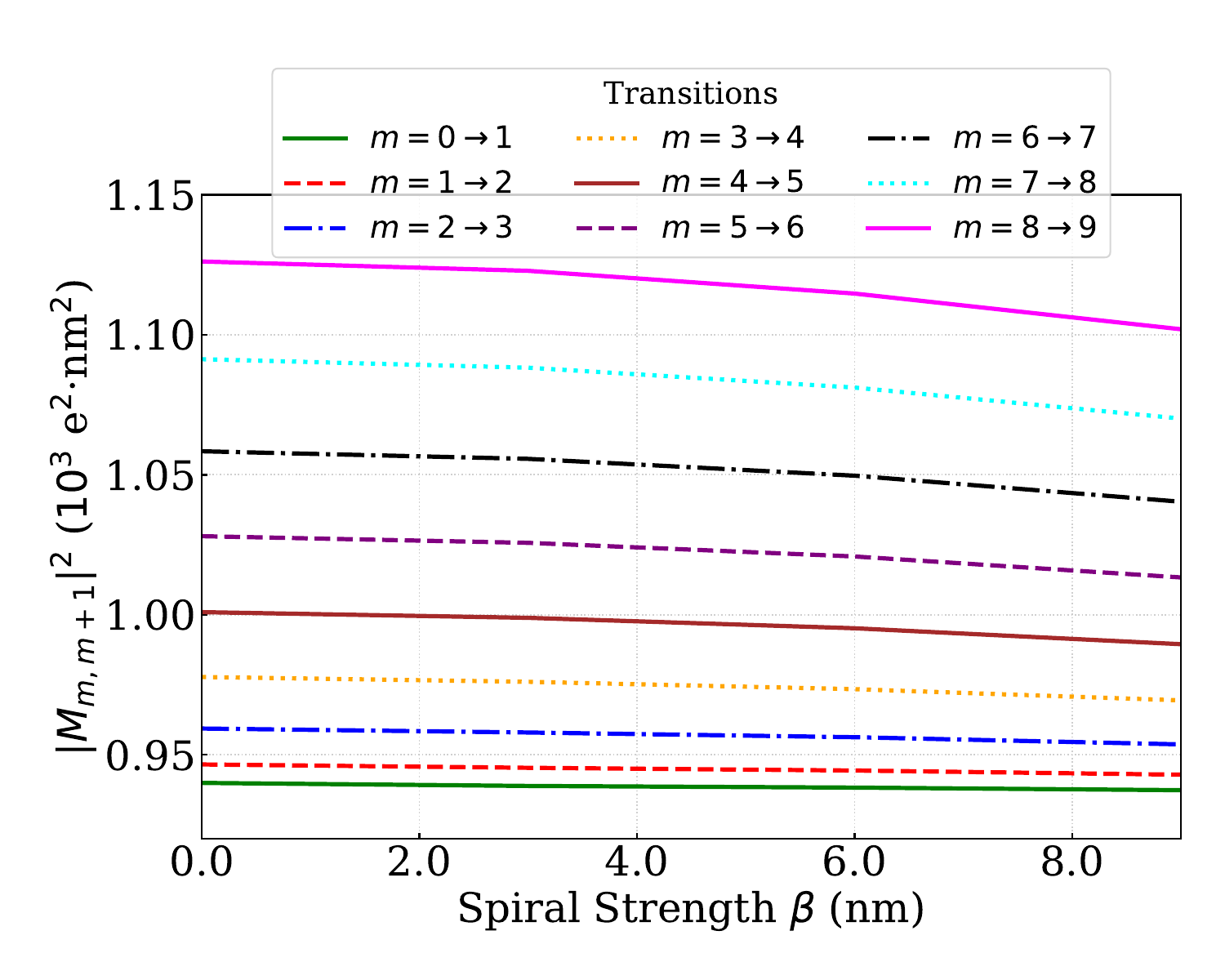}
\includegraphics[width=0.45\textwidth]{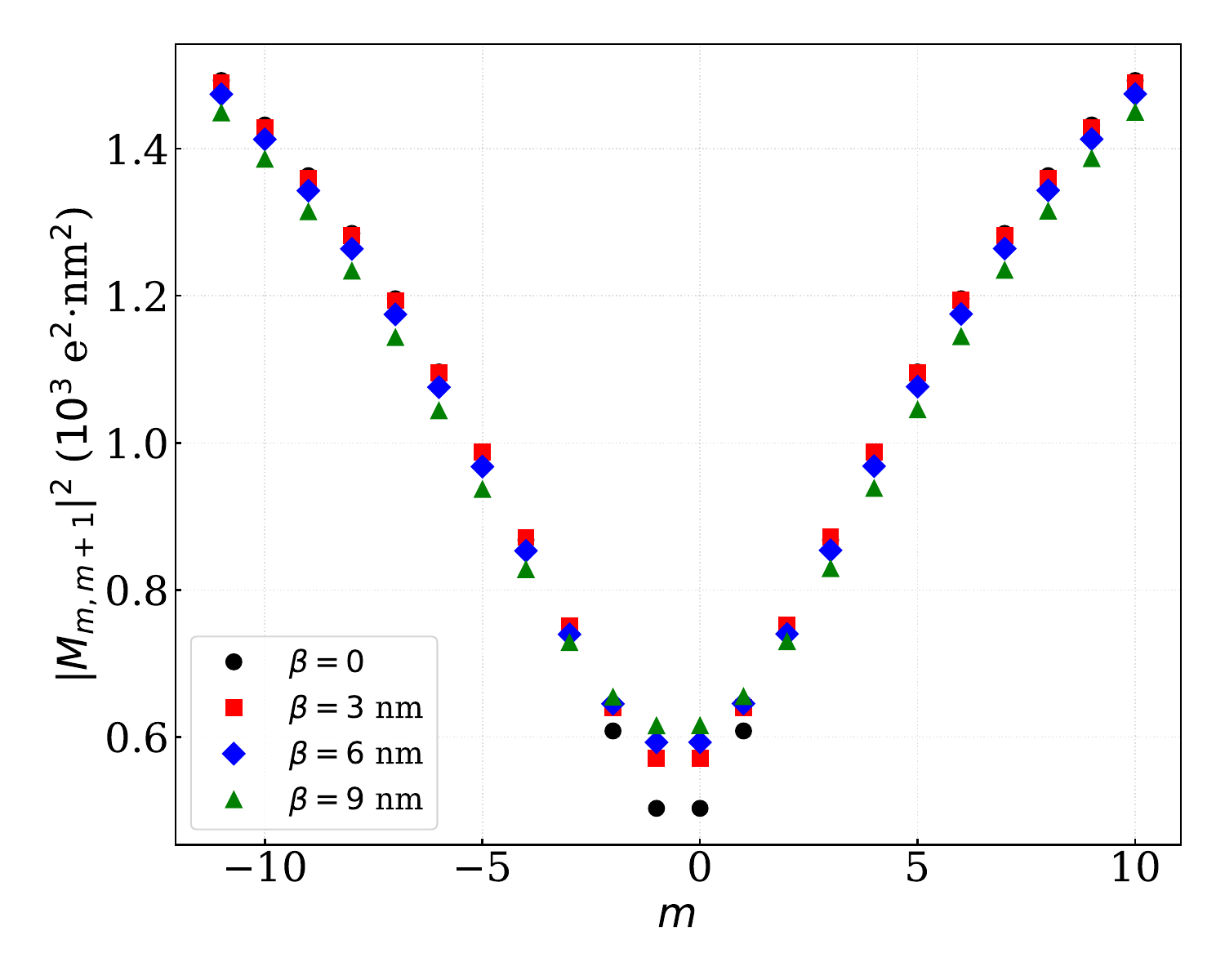}
\caption{(a) Dependence of the squared dipole matrix element $|\mathcal{M}_{m,m+1}|^2$ on the spiral dislocation strength $\beta$ for transitions between consecutive angular momentum states $m \rightarrow m+1$. The values were extracted numerically from the normalized eigenfunctions. (b) Squared dipole matrix element $|\mathcal{M}_{m,m+1}|^2$ as a function of the magnetic quantum number $m$ for different values of the spiral dislocation parameter $\beta = 0$, $3$, $6$, and $9$~nm. The results reveal a clear suppression of the dipole matrix element as $m$ increases, especially for larger values of $\beta$, indicating that the geometric torsion reduces the spatial overlap between initial and final states.}
\label{fig:dipole_matrix}
\end{figure}

Figure~\ref{fig:dipole_matrix}(a) presents the behavior of the squared dipole matrix element $|\mathcal{M}_{m,m+1}|^2$ as a function of the spiral dislocation parameter $\beta$ for different transitions involving consecutive angular momentum states. The results show that, for lower values of $m$, the matrix element remains nearly constant as $\beta$ increases, indicating that the dislocation only weakly affects the transition probability. However, as $m$ increases, the matrix element decreases noticeably with $\beta$. This trend can be attributed to the spatial separation and structural modification of the wavefunctions induced by the spiral geometry. For higher $m$ values, the electron wavefunctions become more localized away from the center of the potential, and the geometric perturbation introduced by the spiral dislocation reduces the spatial overlap between initial and final states. Consequently, the optical transition strength is suppressed for large angular momentum quantum numbers in the presence of a dislocation. This behavior underscores the role of topological defects as tunable mechanisms to control light-matter interactions in quantum-confined systems.

Figure~\ref{fig:dipole_matrix}(b) presents the squared dipole matrix element $|\mathcal{M}_{m,m+1}|^2$ for several angular momentum transitions $m \to m+1$ under different spiral dislocation strengths. For $\beta = 0$, the values of $|\mathcal{M}_{m,m+1}|^2$ decrease smoothly as $m$ increases, reflecting the expected reduction in spatial overlap between radial wavefunctions with higher angular momentum.

However, as the spiral dislocation parameter $\beta$ increases, the suppression becomes more pronounced and nonuniform. This behavior is attributed to the enhanced geometric distortion introduced by the torsional defect, which displaces the wavefunctions further from the origin and diminishes their radial overlap. For higher values of $m$ and $\beta$, the coupling between initial and final states becomes weaker, resulting in a reduced transition probability.

These findings support the idea that spiral dislocations offer a tunable mechanism to modulate light-matter interactions in mesoscopic systems. Specifically, they allow selective suppression of optical transitions involving high angular momentum states, which may be of interest in designing quantum control protocols or tailoring absorption profiles in photonic structures.
\begin{figure}[htb]
\centering
\includegraphics[width=0.44\textwidth]{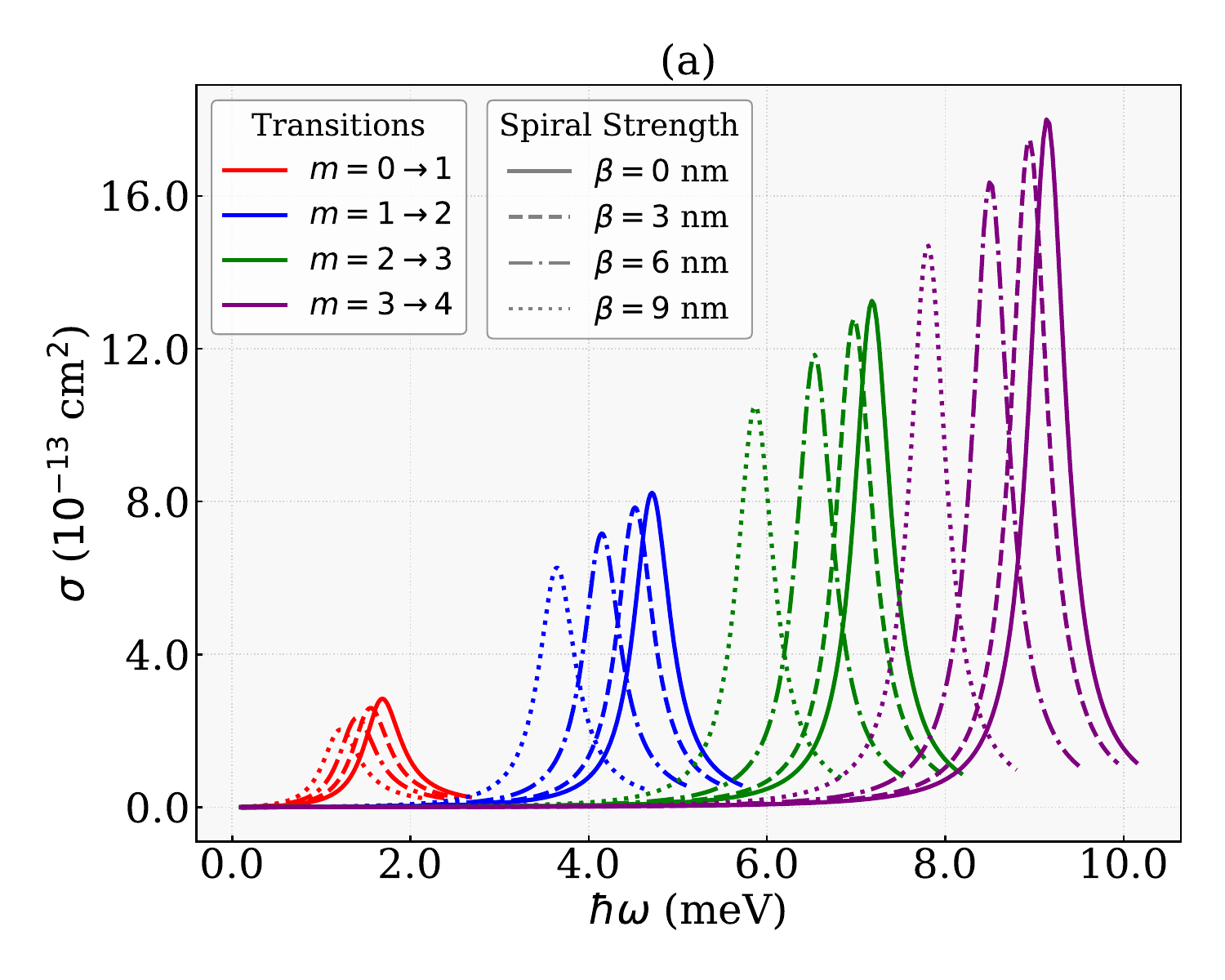}
\includegraphics[width=0.44\textwidth]{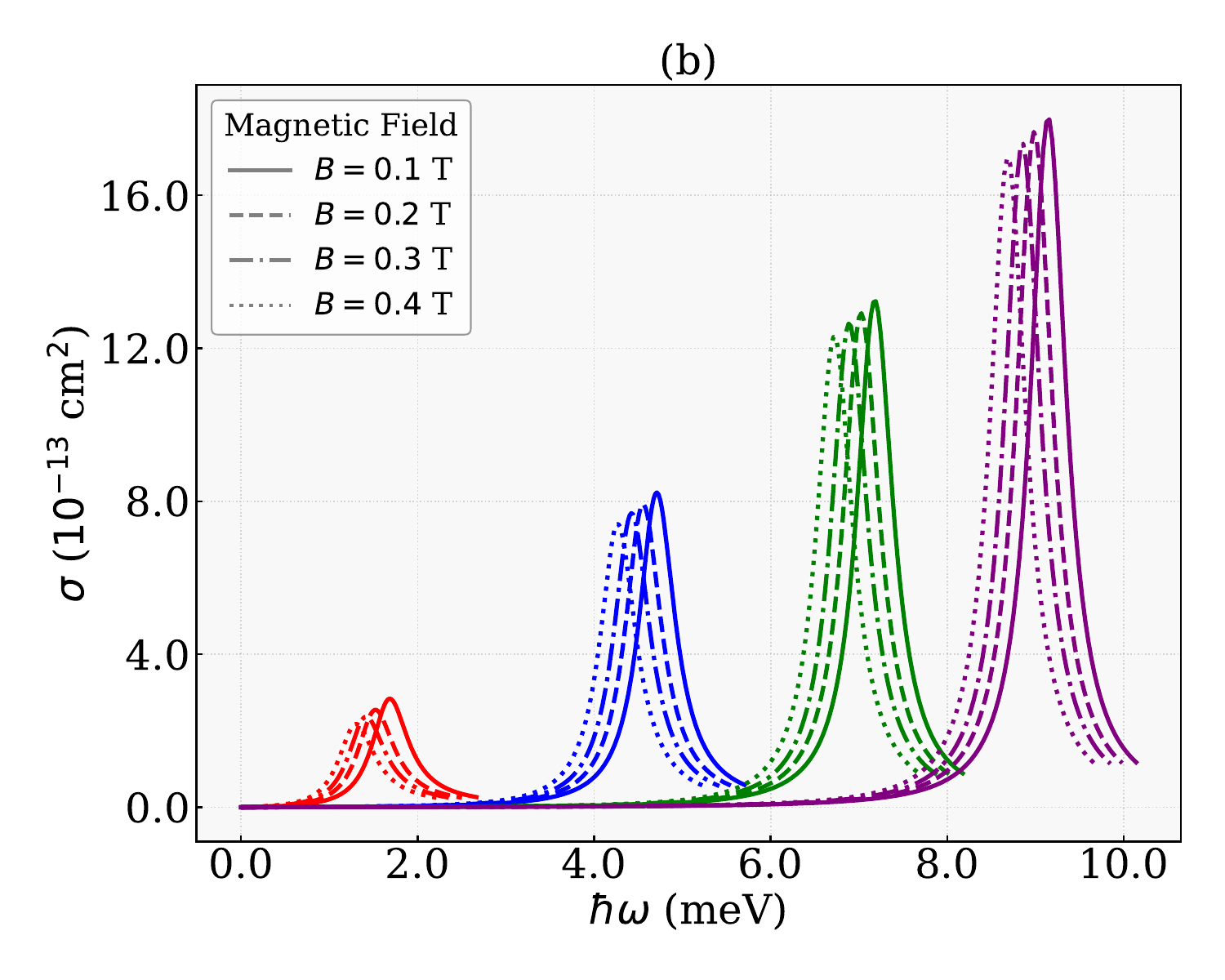}
\caption{Photoionization cross-section as a function of photon energy for several dipole-allowed transitions $m \rightarrow m+1$ in the presence of a spiral dislocation. Each curve corresponds to a specific value of the dislocation parameter $\beta = 0,\, 3,\, 6,\, 9$ nm. As $\beta$ increases, the resonance peaks shift toward higher energies, and the absorption magnitude is significantly enhanced. The effect becomes more pronounced for higher angular momentum transitions, indicating that both $m$ and $\beta$ contribute to modulating the optical response. In (b), we further analyze the role of the magnetic field $B$ for fixed $\beta$, showing that increasing the magnetic field leads to a systematic shift of the resonant peaks and alters the relative intensity of the absorption spectra.}
\label{fig:photoionization_beta}
\end{figure}
\noindent
Figure~\ref{fig:photoionization_beta} illustrates the impact of the spiral dislocation parameter $\beta$ on the photoionization cross-section for several dipole-allowed transitions in a confined quantum system. The results show that increasing $\beta$ consistently enhances the magnitude of the optical absorption response across all transitions analyzed, as a consequence of geometric modifications in the effective potential. Moreover, the resonance peaks shift toward higher photon energies as $\beta$ increases, reflecting an increase in the energy level spacing. The dependence on the magnetic quantum number $m$ is also evident: transitions involving higher values of $m$ lead to broader and more intense absorption profiles, revealing the interplay between topology-induced effects and angular momentum. These findings demonstrate that the spiral dislocation can act as a tunable geometric parameter for tailoring the optical properties of mesoscopic systems.

Figure~\ref{fig:photoionization_beta}(b) complements this analysis by examining the influence of the magnetic field on the photoionization spectra. For a fixed value of the dislocation parameter $\beta$, increasing the magnetic field $B$ leads to a systematic blue shift of the resonant peaks and alters the intensity and shape of the absorption curves. These effects arise from the enhancement of the cyclotron frequency and the associated modifications in the radial confinement, which affect both the energy spectrum and the spatial overlap between initial and final states. Notably, such spectral changes are already observable within a moderate magnetic field regime, highlighting the role of $B$ as a practical external control for tuning the optical response of systems with topological defects.

\begin{table}[htb]
\centering
\caption{Peak transition energies $\Delta E$ (in meV) associated with dipole-allowed transitions $m \to m+1$ for different values of the spiral dislocation parameter $\beta$ and magnetic field $B$. As both $\beta$ and $B$ increase, the transition energy decreases, reflecting a reduction in the energy spacing between bound states.}
\label{tab:transicoes}
\begin{tabular}{cccc}
\toprule
\textbf{Transition} & \boldmath{$\beta$ (nm)} & \boldmath{$B$ (T)} & \boldmath{$\Delta E$ (meV)} \\
\midrule
$0 \to 1$ & 0  & 0.1 & 1.6665 \\
$0 \to 1$ & 3  & 0.2 & 1.4482 \\
$0 \to 1$ & 6  & 0.3 & 1.2227 \\
$0 \to 1$ & 9  & 0.4 & 0.9792 \\
\midrule
$1 \to 2$ & 0  & 0.1 & 4.7028 \\
$1 \to 2$ & 3  & 0.2 & 4.4284 \\
$1 \to 2$ & 6  & 0.3 & 3.9817 \\
$1 \to 2$ & 9  & 0.4 & 3.4202 \\
\midrule
$2 \to 3$ & 0  & 0.1 & 7.1747 \\
$2 \to 3$ & 3  & 0.2 & 6.8862 \\
$2 \to 3$ & 6  & 0.3 & 6.3679 \\
$2 \to 3$ & 9  & 0.4 & 5.6420 \\
\midrule
$3 \to 4$ & 0  & 0.1 & 9.1402 \\
$3 \to 4$ & 3  & 0.2 & 8.8524 \\
$3 \to 4$ & 6  & 0.3 & 8.3379 \\
$3 \to 4$ & 9  & 0.4 & 7.5740 \\
\bottomrule
\end{tabular}
\end{table}
Table~\ref{tab:transicoes} presents the transition energies $\Delta E$ associated with the dominant dipole-allowed processes $m \to m+1$ for different values of the spiral dislocation parameter $\beta$ and magnetic field $B$. The data reveal a clear decreasing trend in $\Delta E$ as both $\beta$ and $B$ increase. This reduction reflects the combined effect of geometric deformations and magnetic confinement on the energy spectrum. 

For a fixed angular momentum $m$, increasing $\beta$ introduces a topological correction to the effective potential, lowering the energy difference between adjacent levels. Simultaneously, increasing $B$ modifies the radial confinement via the cyclotron term, further contributing to the narrowing of the level spacing. These results provide a quantitative basis for understanding the blue shift and intensity variation observed in the photoionization spectra shown in Figure~\ref{fig:photoionization_beta}(b). They also reinforce the notion that both $\beta$ and $B$ serve as external control parameters for tailoring the optical response of the system.

The photoionization cross-section $\sigma(\hbar\omega)$ can be understood as a specific manifestation of optical absorption. In this process, an incident photon with energy $\hbar\omega$ is absorbed by the system, promoting a bound electron to the continuum and effectively ejecting it from the confining potential. This mechanism is conceptually analogous to optical absorption, where energy from light is transferred to the system through dipole-allowed transitions.

The main distinction lies in the nature of the final states. While the conventional absorption coefficient $\alpha(\hbar\omega)$ describes transitions between discrete bound states and is typically used to characterize the attenuation of light in a material, the photoionization cross-section involves transitions from bound states to unbound (continuum) states. Both quantities are governed by the same fundamental principles, such as Fermi's Golden Rule and the dipole matrix element $|\mathcal{M}_{fi}|^2$, and both exhibit resonant behavior characterized by a Lorentzian profile centered at the transition energy.

Therefore, the photoionization cross-section constitutes a meaningful optical response of the system and provides valuable spectroscopic information about its electronic structure. It is particularly relevant in scenarios where light absorption leads to the emission of charge carriers, such as in quantum dots, nanorings, or atomic systems subjected to external electromagnetic fields.

\section{Conclusions \label{conclusion}}

We have studied the quantum and optical behavior of a charged spinless particle confined in a two-dimensional quantum ring subjected to a spiral dislocation and an external magnetic field. The presence of the dislocation was modeled by a torsion-induced metric that introduces off-diagonal geometric terms in the Schrödinger equation without altering the background curvature. By applying the minimal coupling prescription in curved space, we derived a second-order differential equation for the radial component of the wavefunction and transformed it into a canonical Schrödinger-like equation with an energy-dependent effective potential.

Our analysis showed that the spiral dislocation significantly modifies the effective confinement potential, leading to the emergence of asymmetries and multiple potential wells. These geometric distortions alter the energy level spacing, break the $m \leftrightarrow -m$ symmetry, and influence the spatial localization of the wavefunctions.

From the optical perspective, we demonstrated that the spiral dislocation modifies the absorption coefficient, the refractive index variation, and the photoionization cross-section. As the dislocation parameter $\beta$ increases, the resonant peaks shift toward higher photon energies, and the magnitude of the optical response becomes enhanced or suppressed depending on the angular momentum. We also observed a suppression of the dipole matrix elements for higher angular momentum states, confirming that geometric torsion reduces the spatial overlap between initial and final states.

These results highlight the potential of topological defects such as spiral dislocations, as tunable geometric tools for manipulating light-matter interactions in confined quantum systems. The ability to modulate optical properties without altering the external potential opens perspectives for applications in photonic devices, quantum control, and defect-engineered nanostructures.

\section*{Acknowledgement}

This work was partially supported by the Brazilian agencies CAPES, CNPq, and FAPEMA. The author acknowledges the support from the
grants CNPq/306308/2022-3, FAPEMA/UNIVERSAL-06395/22, FAPEMA/APP-12256/22.
This study was partly financed by the Coordenação de
Aperfeiçoamento de Pessoal de Nível Superior - Brazil (CAPES) -
Code 001.

\section*{\label{sec:datar}Data Availability Statement}

Data will be made available on reasonable request.

\bibliographystyle{model1a-num-names}
%\bibliography{mybibliography}

\begin{thebibliography}{41}
\expandafter\ifx\csname natexlab\endcsname\relax\def\natexlab#1{#1}\fi
\providecommand{\bibinfo}[2]{#2}
\ifx\xfnm\relax \def\xfnm[#1]{\unskip,\space#1}\fi
%Type = Inbook
\bibitem[{Srivastava(2001)}]{Srivastava2001}
\bibinfo{author}{A.~M. Srivastava}, \bibinfo{title}{Topological Defects in
  Condensed Matter Physics}, \bibinfo{publisher}{Hindustan Book Agency},
  \bibinfo{address}{Gurgaon}, pp. \bibinfo{pages}{189--237}.
%Type = Article
\bibitem[{Katanaev and Volovich(1992)}]{AoP.1992.216.1}
\bibinfo{author}{M.~Katanaev}, \bibinfo{author}{I.~Volovich},
  \bibinfo{journal}{Annals of Physics} \bibinfo{volume}{216}
  (\bibinfo{year}{1992}) \bibinfo{pages}{1--28}.
%Type = Book
\bibitem[{Monastyrsky(2006)}]{monastyrsky2006topology}
\bibinfo{author}{M.~Monastyrsky}, \bibinfo{title}{Topology in Condensed
  Matter}, Springer Series in Solid-State Sciences,
  \bibinfo{publisher}{Springer Berlin Heidelberg}, \bibinfo{year}{2006}.
%Type = Article
\bibitem[{Mustafa and Guvendi(2025)}]{EPJC.2025.85.34}
\bibinfo{author}{O.~Mustafa}, \bibinfo{author}{A.~Guvendi},
  \bibinfo{journal}{The European Physical Journal C} \bibinfo{volume}{85}
  (\bibinfo{year}{2025}) \bibinfo{pages}{34}.
%Type = Article
\bibitem[{Guvendi et~al.(2021)Guvendi, Zare, and
  Hassanabadi}]{EPJA.2021.57.192}
\bibinfo{author}{A.~Guvendi}, \bibinfo{author}{S.~Zare},
  \bibinfo{author}{H.~Hassanabadi}, \bibinfo{journal}{The European Physical
  Journal A} \bibinfo{volume}{57} (\bibinfo{year}{2021}) \bibinfo{pages}{192}.
%Type = Article
\bibitem[{Moreira and Ahmed(2024)}]{IJP.2024.98.4827}
\bibinfo{author}{A.~R.~P. Moreira}, \bibinfo{author}{F.~Ahmed},
  \bibinfo{journal}{Indian Journal of Physics} \bibinfo{volume}{98}
  (\bibinfo{year}{2024}) \bibinfo{pages}{4827--4834}.
%Type = Article
\bibitem[{Maia and Bakke(2020)}]{AoP.2020.419.168229}
\bibinfo{author}{A.~Maia}, \bibinfo{author}{K.~Bakke}, \bibinfo{journal}{Annals
  of Physics} \bibinfo{volume}{419} (\bibinfo{year}{2020})
  \bibinfo{pages}{168229}.
%Type = Article
\bibitem[{Maia and Bakke(2018)}]{PB.2018.531.213}
\bibinfo{author}{A.~Maia}, \bibinfo{author}{K.~Bakke},
  \bibinfo{journal}{Physica B: Condensed Matter} \bibinfo{volume}{531}
  (\bibinfo{year}{2018}) \bibinfo{pages}{213--215}.
%Type = Article
\bibitem[{Zare et~al.(2022)Zare, Hassanabadi, and Guvendi}]{EPJP.2022.137.589}
\bibinfo{author}{S.~Zare}, \bibinfo{author}{H.~Hassanabadi},
  \bibinfo{author}{A.~Guvendi}, \bibinfo{journal}{The European Physical Journal
  Plus} \bibinfo{volume}{137} (\bibinfo{year}{2022}) \bibinfo{pages}{589}.
%Type = Article
\bibitem[{Filgueiras and Silva(2015)}]{PLA.2015.379.2110}
\bibinfo{author}{C.~Filgueiras}, \bibinfo{author}{E.~O. Silva},
  \bibinfo{journal}{Physics Letters A} \bibinfo{volume}{379}
  (\bibinfo{year}{2015}) \bibinfo{pages}{2110--2115}.
%Type = Article
\bibitem[{Furtado and Moraes(1999)}]{EPL.1999.45.279}
\bibinfo{author}{C.~Furtado}, \bibinfo{author}{F.~Moraes},
  \bibinfo{journal}{Europhysics Letters} \bibinfo{volume}{45}
  (\bibinfo{year}{1999}) \bibinfo{pages}{279}.
%Type = Article
\bibitem[{Filgueiras and de~Oliveira(2011)}]{AdP.2011.523.898}
\bibinfo{author}{C.~Filgueiras}, \bibinfo{author}{B.~de~Oliveira},
  \bibinfo{journal}{Annalen der Physik} \bibinfo{volume}{523}
  (\bibinfo{year}{2011}) \bibinfo{pages}{898--909}.
%Type = Article
\bibitem[{Filgueiras et~al.(2016)Filgueiras, Rojas, Aciole, and
  Silva}]{PLA.2016.380.3847}
\bibinfo{author}{C.~Filgueiras}, \bibinfo{author}{M.~Rojas},
  \bibinfo{author}{G.~Aciole}, \bibinfo{author}{E.~O. Silva},
  \bibinfo{journal}{Physics Letters A} \bibinfo{volume}{380}
  (\bibinfo{year}{2016}) \bibinfo{pages}{3847--3853}.
%Type = Article
\bibitem[{Budagovsky et~al.(2015)Budagovsky, Zolot'ko, Korshunov, Smayev,
  Shvetsov, and Barnik}]{OS.2015.119.280}
\bibinfo{author}{I.~A. Budagovsky}, \bibinfo{author}{A.~S. Zolot'ko},
  \bibinfo{author}{D.~L. Korshunov}, \bibinfo{author}{M.~P. Smayev},
  \bibinfo{author}{S.~A. Shvetsov}, \bibinfo{author}{M.~I. Barnik},
  \bibinfo{journal}{Optics and Spectroscopy} \bibinfo{volume}{119}
  (\bibinfo{year}{2015}) \bibinfo{pages}{280--285}.
%Type = Article
\bibitem[{Maia and Bakke(2019)}]{EPJC.2019.79.551}
\bibinfo{author}{A.~V. D.~M. Maia}, \bibinfo{author}{K.~Bakke},
  \bibinfo{journal}{The European Physical Journal C} \bibinfo{volume}{79}
  (\bibinfo{year}{2019}) \bibinfo{pages}{551}.
%Type = Article
\bibitem[{Hassanabadi et~al.(2021)Hassanabadi, Zare, L\"{u}tf\"{u}o\u{g}lu,
  K\v{r}\'{\i}\v{z}, and Hassanabadi}]{IJMPA.2021.36.2150100}
\bibinfo{author}{S.~Hassanabadi}, \bibinfo{author}{S.~Zare},
  \bibinfo{author}{B.~C. L\"{u}tf\"{u}o\u{g}lu},
  \bibinfo{author}{J.~K\v{r}\'{\i}\v{z}}, \bibinfo{author}{H.~Hassanabadi},
  \bibinfo{journal}{International Journal of Modern Physics A}
  \bibinfo{volume}{36} (\bibinfo{year}{2021}) \bibinfo{pages}{2150100}.
%Type = Article
\bibitem[{{da Silva} and Bakke(2020)}]{AoP.2020.421.168277}
\bibinfo{author}{W.~{da Silva}}, \bibinfo{author}{K.~Bakke},
  \bibinfo{journal}{Annals of Physics} \bibinfo{volume}{421}
  (\bibinfo{year}{2020}) \bibinfo{pages}{168277}.
%Type = Article
\bibitem[{Valanis and Panoskaltsis(2005)}]{AM.2005.175.77}
\bibinfo{author}{K.~C. Valanis}, \bibinfo{author}{V.~P. Panoskaltsis},
  \bibinfo{journal}{Acta Mechanica} \bibinfo{volume}{175}
  (\bibinfo{year}{2005}) \bibinfo{pages}{77--103}.
%Type = Book
\bibitem[{Landau and Lifshitz(1991)}]{landau1991quantum}
\bibinfo{author}{L.~D. Landau}, \bibinfo{author}{E.~M. Lifshitz},
  \bibinfo{title}{Quantum Mechanics: Non-Relativistic Theory},
  \bibinfo{publisher}{Butterworth-Heinemann}, \bibinfo{year}{1991}.
%Type = Book
\bibitem[{Stone(1992)}]{stone1992quantum}
\bibinfo{author}{M.~Stone}, \bibinfo{title}{Quantum Hall Effect},
  \bibinfo{publisher}{World Scientific}, \bibinfo{year}{1992}.
%Type = Article
\bibitem[{Tan and Inkson(1996)}]{SST.1996.11.1635}
\bibinfo{author}{W.~Tan}, \bibinfo{author}{J.~Inkson},
  \bibinfo{journal}{Semiconductor science and technology} \bibinfo{volume}{11}
  (\bibinfo{year}{1996}) \bibinfo{pages}{1635}.
%Type = Article
\bibitem[{Pereira et~al.(2024{\natexlab{a}})Pereira, dos S~Azevedo, Pereira,
  and Silva}]{CTP.2024.76.105701}
\bibinfo{author}{C.~M.~O. Pereira}, \bibinfo{author}{F.~dos S~Azevedo},
  \bibinfo{author}{L.~F.~C. Pereira}, \bibinfo{author}{E.~O. Silva},
  \bibinfo{journal}{Communications in Theoretical Physics} \bibinfo{volume}{76}
  (\bibinfo{year}{2024}{\natexlab{a}}) \bibinfo{pages}{105701}.
%Type = Article
\bibitem[{Pereira et~al.(2024{\natexlab{b}})Pereira, Azevedo, and
  Silva}]{QR.2024.6.677}
\bibinfo{author}{C.~M.~O. Pereira}, \bibinfo{author}{F.~d.~S. Azevedo},
  \bibinfo{author}{E.~O. Silva}, \bibinfo{journal}{Quantum Reports}
  \bibinfo{volume}{6} (\bibinfo{year}{2024}{\natexlab{b}})
  \bibinfo{pages}{677--705}.
%Type = Article
\bibitem[{Şahin(2008)}]{PRB.2008.77.045317}
\bibinfo{author}{M.~Şahin}, \bibinfo{journal}{Physical Review B}
  \bibinfo{volume}{77} (\bibinfo{year}{2008}) \bibinfo{pages}{045317}.
%Type = Article
\bibitem[{Sali et~al.(1999)Sali, Duque, and Mora-Ramos}]{PSSB.1999.211.611}
\bibinfo{author}{A.~Sali}, \bibinfo{author}{C.~A. Duque},
  \bibinfo{author}{M.~E. Mora-Ramos}, \bibinfo{journal}{phys. stat. sol. (b)}
  \bibinfo{volume}{211} (\bibinfo{year}{1999}) \bibinfo{pages}{661--670}.
%Type = Article
\bibitem[{Hahn et~al.(2025)Hahn, Duque, and Mora-Ramos}]{PLA.2025.130226}
\bibinfo{author}{R.~Hahn}, \bibinfo{author}{C.~Duque},
  \bibinfo{author}{M.~Mora-Ramos}, \bibinfo{journal}{Physics Letters A}
  \bibinfo{volume}{534} (\bibinfo{year}{2025}) \bibinfo{pages}{130226}.
%Type = Article
\bibitem[{Duque et~al.(2012)Duque, Mora-Ramos, and Duque}]{AdP.2012.524.327}
\bibinfo{author}{C.~Duque}, \bibinfo{author}{M.~Mora-Ramos},
  \bibinfo{author}{C.~Duque}, \bibinfo{journal}{Annalen der Physik}
  \bibinfo{volume}{524} (\bibinfo{year}{2012}) \bibinfo{pages}{327--337}.
%Type = Article
\bibitem[{Suvajit~Pal and Duque(2019)}]{PM.2019.99.2457}
\bibinfo{author}{M.~G. Suvajit~Pal}, \bibinfo{author}{C.~A. Duque},
  \bibinfo{journal}{Philosophical Magazine} \bibinfo{volume}{99}
  (\bibinfo{year}{2019}) \bibinfo{pages}{2457--2486}.
%Type = Article
\bibitem[{Duan et~al.(2022)Duan, Li, Chang, and Zhao}]{Optik.2022.261.169187}
\bibinfo{author}{Y.~Duan}, \bibinfo{author}{X.~Li}, \bibinfo{author}{C.~Chang},
  \bibinfo{author}{Z.~Zhao}, \bibinfo{journal}{Optik} \bibinfo{volume}{261}
  (\bibinfo{year}{2022}) \bibinfo{pages}{169187}.
%Type = Article
\bibitem[{Máthé et~al.(2021)Máthé, Onyenegecha, Farcaş,
  Pioraş-Ţimbolmaş, Solaimani, and Hassanabadi}]{PLA.2021.397.127262}
\bibinfo{author}{L.~Máthé}, \bibinfo{author}{C.~Onyenegecha},
  \bibinfo{author}{A.-A. Farcaş}, \bibinfo{author}{L.-M. Pioraş-Ţimbolmaş},
  \bibinfo{author}{M.~Solaimani}, \bibinfo{author}{H.~Hassanabadi},
  \bibinfo{journal}{Physics Letters A} \bibinfo{volume}{397}
  (\bibinfo{year}{2021}) \bibinfo{pages}{127262}.
%Type = Article
\bibitem[{Candemir and Özdemir(2023)}]{PLA.2023.492.129226}
\bibinfo{author}{N.~Candemir}, \bibinfo{author}{A.~Özdemir},
  \bibinfo{journal}{Physics Letters A} \bibinfo{volume}{492}
  (\bibinfo{year}{2023}) \bibinfo{pages}{129226}.
%Type = Article
\bibitem[{Talwar et~al.(2022)Talwar, Lumb, and Prasad}]{EPJP.2022.137.175}
\bibinfo{author}{S.~L. Talwar}, \bibinfo{author}{S.~Lumb},
  \bibinfo{author}{V.~Prasad}, \bibinfo{journal}{The European Physical Journal
  Plus} \bibinfo{volume}{137} (\bibinfo{year}{2022}) \bibinfo{pages}{175}.
%Type = Article
\bibitem[{Liang et~al.(2011)Liang, Xie, and Shen}]{OC.2011.284.5818}
\bibinfo{author}{S.~Liang}, \bibinfo{author}{W.~Xie},
  \bibinfo{author}{H.~Shen}, \bibinfo{journal}{Optics Communications}
  \bibinfo{volume}{284} (\bibinfo{year}{2011}) \bibinfo{pages}{5818--5828}.
%Type = Article
\bibitem[{Aydin et~al.(2021)Aydin, Sari, Kasapoglu, Sakiroglu, and
  S{\"o}kmen}]{EPJP.2021136.832}
\bibinfo{author}{F.~Aydin}, \bibinfo{author}{H.~Sari},
  \bibinfo{author}{E.~Kasapoglu}, \bibinfo{author}{S.~Sakiroglu},
  \bibinfo{author}{I.~S{\"o}kmen}, \bibinfo{journal}{The European Physical
  Journal Plus} \bibinfo{volume}{136} (\bibinfo{year}{2021})
  \bibinfo{pages}{832}.
%Type = Article
\bibitem[{Yuan et~al.(2018)Yuan, Wang, Xiong, Chen, Zhang, and
  Zhao}]{EPJP.2018.133.395}
\bibinfo{author}{J.-H. Yuan}, \bibinfo{author}{L.-L. Wang},
  \bibinfo{author}{Z.-Y. Xiong}, \bibinfo{author}{N.~Chen},
  \bibinfo{author}{Z.-H. Zhang}, \bibinfo{author}{Y.-X. Zhao},
  \bibinfo{journal}{The European Physical Journal Plus} \bibinfo{volume}{133}
  (\bibinfo{year}{2018}) \bibinfo{pages}{395}.
%Type = Article
\bibitem[{Baig(2022)}]{Atoms.2022.10.39}
\bibinfo{author}{M.~A. Baig}, \bibinfo{journal}{Atoms} \bibinfo{volume}{10}
  (\bibinfo{year}{2022}) \bibinfo{pages}{39}.
%Type = Article
\bibitem[{Azmi et~al.(2024)Azmi, Amri, Nithiananthi, Jaouane, El-Bakkari, Sali,
  Ed-Dahmouny, Fakkahi, and Arraoui}]{PB.2024.677.415717}
\bibinfo{author}{H.~Azmi}, \bibinfo{author}{N.~Amri},
  \bibinfo{author}{P.~Nithiananthi}, \bibinfo{author}{M.~Jaouane},
  \bibinfo{author}{K.~El-Bakkari}, \bibinfo{author}{A.~Sali},
  \bibinfo{author}{A.~Ed-Dahmouny}, \bibinfo{author}{A.~Fakkahi},
  \bibinfo{author}{R.~Arraoui}, \bibinfo{journal}{Physica B: Condensed Matter}
  \bibinfo{volume}{677} (\bibinfo{year}{2024}) \bibinfo{pages}{415717}.
%Type = Article
\bibitem[{Zeiri et~al.(2025)Zeiri, Baser, Jahromi, Yahyaoui, Ed-Dahmouny,
  Sfina, Duque, and Said}]{OLT.2025.182.111822}
\bibinfo{author}{N.~Zeiri}, \bibinfo{author}{P.~Baser}, \bibinfo{author}{H.~D.
  Jahromi}, \bibinfo{author}{N.~Yahyaoui}, \bibinfo{author}{A.~Ed-Dahmouny},
  \bibinfo{author}{N.~Sfina}, \bibinfo{author}{C.~Duque},
  \bibinfo{author}{M.~Said}, \bibinfo{journal}{Optics \& Laser Technology}
  \bibinfo{volume}{182} (\bibinfo{year}{2025}) \bibinfo{pages}{111822}.
%Type = Article
\bibitem[{Shi and Yan(2023)}]{PLA.2023.466.128725}
\bibinfo{author}{L.~Shi}, \bibinfo{author}{Z.~Yan}, \bibinfo{journal}{Physics
  Letters A} \bibinfo{volume}{466} (\bibinfo{year}{2023})
  \bibinfo{pages}{128725}.
%Type = Article
\bibitem[{Azmi et~al.(2025)Azmi, El-Bakkari, Jaouane, Fakkahi, Arraoui,
  Ed-Dahmouny, Sali, Amri, {El Ghazi}, and Duque}]{PB.2025.696.416647}
\bibinfo{author}{H.~Azmi}, \bibinfo{author}{K.~El-Bakkari},
  \bibinfo{author}{M.~Jaouane}, \bibinfo{author}{A.~Fakkahi},
  \bibinfo{author}{R.~Arraoui}, \bibinfo{author}{A.~Ed-Dahmouny},
  \bibinfo{author}{A.~Sali}, \bibinfo{author}{N.~Amri}, \bibinfo{author}{H.~{El
  Ghazi}}, \bibinfo{author}{C.~Duque}, \bibinfo{journal}{Physica B: Condensed
  Matter} \bibinfo{volume}{696} (\bibinfo{year}{2025}) \bibinfo{pages}{416647}.
%Type = Article
\bibitem[{Nguepi et~al.(2025)Nguepi, Tshipa, and
  Keolopile}]{PS.2025.100.055914}
\bibinfo{author}{F.~G. Nguepi}, \bibinfo{author}{M.~Tshipa},
  \bibinfo{author}{Z.~G. Keolopile}, \bibinfo{journal}{Physica Scripta}
  \bibinfo{volume}{100} (\bibinfo{year}{2025}) \bibinfo{pages}{055914}.

\end{thebibliography}

\end{document}